\documentclass[preprint,11pt]{elsarticle}

\usepackage{graphicx}

\usepackage{amssymb}
\usepackage{amsthm}
\usepackage{amsmath}
\usepackage{color}
\usepackage{lineno}

\biboptions{comma,sort&compress}

\journal{Physica D}

\begin{document}

\begin{frontmatter}


\title{Bifurcations and strange nonchaotic attractors in a phase oscillator model of glacial-interglacial cycles
}

\author[label1]{Takahito Mitsui\corref{cor1}}
\cortext[cor1]{Corresponding author.}
\ead{takahito321@gmail.com}
\author[label1]{Michel Crucifix}
\author[label2]{Kazuyuki Aihara}
\address[label1]{Georges Lema\^\i tre Centre for Earth and Climate Research, Earth and Life Institute, Universit\'e catholique de Louvain, Louvain-la-Neuve, Belgium}
\address[label2]{Institute of Industrial Science, University of Tokyo, 4-6-1 Komaba, Meguro-ku, Tokyo 153-8505, Japan}

\begin{abstract}
Glacial-interglacial cycles are large variations in continental ice mass and greenhouse gases, which have dominated climate variability over the Quaternary. 
The dominant periodicity of the cycles is $\sim $40~kyr before the so-called middle Pleistocene transition between $\sim$1.2 and $\sim$0.7 Myr ago, and it is $\sim $100~kyr after the transition.
In this paper, the dynamics of glacial-interglacial cycles are investigated using a phase oscillator model forced by the time-varying incoming solar radiation (insolation). We analyze the bifurcations of the system and show that strange nonchaotic attractors appear through nonsmooth saddle-node bifurcations of tori. 
The bifurcation analysis indicates that mode-locking is likely to occur for the 41~kyr glacial cycles but not likely for the 100~kyr glacial cycles. The sequence of mode-locked 41~kyr cycles is robust to small parameter changes. However, the sequence of 100~kyr glacial cycles can be sensitive to parameter changes when the system has a strange nonchaotic attractor.
\end{abstract}


\end{frontmatter}


\section{Introduction}\label{Sec1}
The glacial-interglacial cycles or simply glacial cycles are alternations between cold (glacial) and warm (interglacial) periods, which occurred over the last 3 million years (Myr). They are characterized by large fluctuations in the continental ice volume, which can be estimated, among others, from the oxygen isotope ratio $\delta ^{18}O$ of benthic foraminifera sampled in deep-sea cores (see Fig.~\ref{fig:MPT_bb}) \cite{rf:Pail2015,rf:Imbr1992,rf:Imbr1993,rf:Rudd2003,rf:Lisi2005}. A large value of $\delta ^{18}O$ indicates a large volume of ice sheets. The dominant period of the glacial cycles is $\sim $40~kyr before $\sim $1.2~Myr ago and $\sim $100~kyr after $\sim $0.7~Myr ago. This frequency change accompanying an increase amplitude is called the middle-Pleistocene transition (MPT) (see \cite{rf:Clar2006} and references therein).    

There is abundant empirical evidence \cite{rf:Pail2015,rf:Imbr1992,rf:Imbr1993,rf:Rudd2003,rf:Lisi2005,rf:Clar2006,rf:Hays1976,rf:Lisi2010,rf:Nie2008,rf:Rial2013} that glacial cycles are controlled by the changes in the parameters of Earth's orbit (eccentricity $e$ and longitude of perihelion $\varpi$) and its obliquity $\varepsilon$ \cite{rf:Berg1978}. Milankovitch's theory provides a physical mechanism for this control \cite{rf:Mila1941}: the orbital and obliquity changes determine the seasonal and spatial distributions of insolation (this is called the {\it astronomical forcing}), and, specifically, northern hemisphere summer insolation controls the interannual accumulation of snow and the subsequent growth of ice sheets. Milankovitch's mechanism is nowadays viewed as realistic, but it is considered to be only one element of a complex nonlinear dynamical process. For example, the glacial cycles of the last 0.7~Myr have a dominant spectral power near 100~kyr, while the astronomical forcing has negligible power near 100~kyr (so called the {\it 100-kyr cycle problem} \cite{rf:Imbr1993}). Thus, some nonlinear transformation mechanism must exist from astronomical forcing to glacial cycles \cite{rf:Hays1976}.

The glacial cycles have been investigated using various models ranging from simple conceptual models to complex coupled general circulation models (for example, \cite{rf:Abe2013}), between which there are Earth Systems Models of Intermediate Complexity (EMICs) \cite{rf:Clau2002,rf:Gall1992,rf:Gano2010}. Among others, conceptual models provide theoretical frameworks to identify the dynamics of glacial cycles (for example, see \cite{rf:Pail2015,rf:Cruc2012a,rf:Ghil1987,rf:Salt2001,rf:Weer1976,rf:Oerl1982,rf:Tzip2006,rf:Hogg2008,rf:Cruc2012b,rf:Saed2013,rf:Mits2014,rf:Cruc2013,rf:Daru2014,rf:Ashk2004} and references therein).
They are presented under the form of low-dimensional dynamical systems forced by a measure of northern hemisphere insolation or, more generally, by a linear combination of $e\sin \varpi$ (the {\it climatic precession}) and obliquity $\varepsilon$. 
Since these quantities are well-approximated as quasiperiodic functions of time over the past several million years \cite{rf:Berg1978}, the models of glacial cycles may be viewed as quasiperiodically forced dynamical systems.

Let $\mathbb{T}^N=\mathbb{R}^N/(2\pi \mathbb{Z})^N$ be an $N$-dimensional torus.
Quasiperiodically forced dynamical systems can be represented in a skew-product form:
\begin{equation}
\left\{
\begin{split} 
\dot{\mathbf{\theta }}&=\mathbf{\omega },\ \ \ \ \ \ \ \ \theta \in \mathbb{T}^N, \\
\dot{\mathbf{x}}&=\mathbf{f}\mathbf{(x},\mathbf{\theta }), \ \mathbf{x}\in \mathbb{R}^M,
\end{split} \label{eq:general}
\right.
\end{equation}
where $\mathbf{\theta }=(\theta _1, \theta _2, ..., \theta _N)^\mathrm{T}$ is the phase of the drive system, $\mathbf{x}=(x _1, x _2, ..., x _M)^\mathrm{T}$ is the state of the response system,  
$\mathbf{f}\mathbf{(x},\mathbf{\theta })$ is a periodic function in each phase $\theta _i$, 
and $\mathbf{\omega }=(\omega _1, \omega _2, ..., \omega _N)^\mathrm{T}$ is a vector of incommensurate frequencies such that $k_1\omega _1+k_2\omega _2+\dots +k_N\omega _N= 0$ does not hold for any set of integers, $k_1$, $k_2$, ..., $k_N$, except for 
the trivial solution $k_1=k_2=\dots =k_N= 0$.
In the models of glacial cycles, 
$\theta (t)$ corresponds to the phase of the astronomical forcing, and $\mathbf{x}(t)$ corresponds to the climate state.

Quasiperiodically forced systems can exhibit intermediate dynamics between quasiperiodicity and chaos, so-called {\it strange nonchaotic attractors} (SNAs) \cite{rf:Greb1984,rf:Kane1984}. An SNA is a geometrically strange attractor for which typical Lyapunov exponents are nonpositive \cite{rf:Greb1984} (see \cite{rf:Pras2001,rf:Feud2006} for comprehensive reviews). 
The system~(\ref{eq:general}) is characterized by ($N+M$) Lyapunov exponents. Among them, $N$ Lyapunov exponents are trivially zero. They correspond to the phase equations of the drive system. When the system has an SNA, the nontrivial largest Lyapunov exponent is negative in general. Thus, under a common quasiperiodic forcing, trajectories $\mathbf{x}(t)$, which start from different initial conditions $\mathbf{x}(0)$, approach (or synchronize to) a unique, or one of a finite number of possible  trajectories as time elapses (this is a {\it synchronizing property}) \cite{rf:Rama1997,rf:Ueno2013}. However, related to the strange geometry of SNAs, trajectories $\mathbf{x}(t)$ have nonexponential sensitivity on initial phases $\theta (0)$ \cite{rf:Piko1995,rf:Feud1995,rf:Glen2006,rf:Mits2011} or on parameter values of the systems \cite{rf:Nish1996}. SNAs have been observed in many laboratory experiments \cite{rf:Ditt1990,rf:Zhou1992,rf:Tang1997,rf:Tham2006,rf:Ruiz2007,rf:Mits2012,rf:Ueno2013} but very rarely in nature so far. Linder {\it et al.} show that the brightness changes of some RRc Lyrae stars have nonchaotic and fractal properties of SNAs \cite{rf:Lind2015}. 

Recently, the authors of this paper showed that several models of glacial cycles exhibit SNAs \cite{rf:Mits2014,rf:Cruc2013}.
This means that the relationship from the phase of astronomical forcing $\theta $ to the climate state $\mathbf{x}$ is represented by a geometrically strange set. When a model of glacial cycles exhibits an SNA, the sequences of glacial cycles synchronize under the same astronomical forcing, but simultaneously they can be sensitive to parameter changes. Ivashchenko {\it et al.} showed that the autocorrelation function of the benthic isotopic data has self-simility characteristic of SNAs \cite{rf:Ivas2014}.

However, so far, the birth mechanism of SNAs has not been elucidated in the models of glacial cycles. In this paper, we introduce a phase oscillator model of glacial cycles, whose bifurcation analysis is easier than higher-dimensional models, and show the birth mechanism of SNAs in this model. Based on the bifurcation analysis of the phase model, we suggest that the 41-kyr cycles, typical before the MPT, are little sensitive to the parameters. By contrast the sequence of 100-kyr cycles, which characterises climate after the MPT, can be strongly sensitive to small parameter changes. 

The remainder of this article is organized as follows. In Section 2, a phase
oscillator model for glacial cycles is introduced. In Section 3, the bifurcations exhibited by the model are analyzed for two types of external forcing: an ideal
two-frequency quasiperiodic forcing and the astronomical forcing. In Section
4, we discuss parameter sensitivity of the phase model, and the MPT. Section~5 summarizes this article. In Appendix, we characterize the strangeness of attractors using the phase sensitivity exponent \cite{rf:Piko1995}.

\section{Model}\label{Sec2}
Consistently with the above discussion,
the astronomical forcing $F_A(t)$ is calculated by the following formula \cite{rf:Saed2013}:
\begin{equation} 
F_A(t)=\frac{1}{a}\sum _{i=1}^{35} (s_i\sin \omega _it+c_i\cos \omega _it), \label{eq:ins}
\end{equation}
where we set $a=23.58$ W/m$^2$ so that $F_A(t)$ has unit variance.  
The frequencies $\omega _i$ and coefficients, $s_i$ and $c_i$, are based on \cite{rf:Saed2013}, and
they are listed in \cite{rf:Mits2014} in descending order of power $s_i^2+c_i^2$.
The first three periods correspond to the climatic precession: $2\pi /\omega _1\approx 23.7$~kyr, $2\pi /\omega _2\approx 22.4$~kyr, and $2\pi /\omega _3\approx 19.0$~kyr. The fourth period $2\pi /\omega _4\approx 41.0$~kyr corresponds to the obliquity change \cite{rf:Berg1978}. Unless mentioned otherwise, we use the time unit of 10~kyr and the angular unit of radian for the variables and parameters of the model.    

\begin{figure}[t]
\begin{center}
\includegraphics[scale=0.6,angle=0]{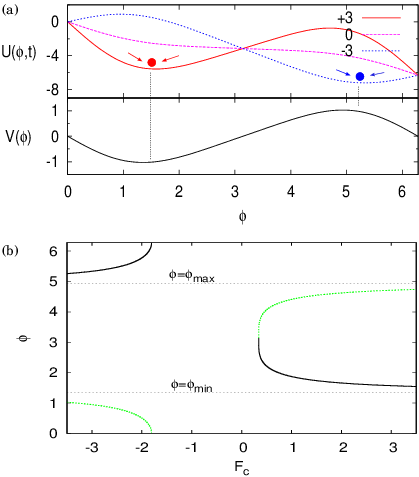}
\caption{(Color) (a) Ice volume $V(\phi )$ (black solid line) and potential $U(\phi ,t)$ as functions of $\phi $ for $F(t)=3$ (red solid line), $F(t)=0$ (magenta dashed line), and $F(t)=-3$ (blue dotted line).
$\alpha =1.0$, $\beta =1.0006$, $\gamma =1.0$, and $\delta =0.24$. (b) Bifurcation diagram of Eq.~(\ref{eq:qp}) for constant forcing $F(t)=F_c$.
The positions of stable nodes and saddles are presented by black solid lines and green dotted lines, respectively. The horizontal lines are drawn at the ice volume maximum $\phi =\phi _{\max}$ and minimum $\phi =\phi _{\min}$. 
} \label{fig:potential_paper}
\end{center}
\end{figure}

Glacial cycles are described here by a forced oscillator, as suggested by some recent analyses \cite{rf:Lisi2010,rf:Nie2008,rf:Rial2013}.
Further, the global ice volume $V(\phi )$ is modeled as a $2\pi$-periodic function of the phase of the oscillator $\phi (t)\in \mathbb{T}^1=\mathbb{R}^1/(2\pi \mathbb{Z})$, i.e., $V(\phi )=V(\phi +2\pi )$.
We express the ice volume $V(\phi )$ as a Fourier sine series up to the second harmonic: 
\begin{equation}
V(\phi )=-\sin \phi -\frac{\delta }{2}\sin 2\phi ,\label{eq:V}
\end{equation}
where $\delta $ is the modification parameter.
For $|\delta| <1$, the ice volume $V(\phi)$ has a unique minimum at $\phi _{\min}=\cos ^{-1}[(\sqrt{1+8\delta ^2}-1)/(4\delta)]$ and a maximum at $\phi _{\max}=2\pi -\phi _{\min}$, as shown in Fig.~\ref{fig:potential_paper}(a).

In general, the phase motion of a limit-cycle oscillator under a weak forcing $\mathbf{p}(t)$ is described by
$\dot{\phi}=\beta +\mathbf{Z}(\phi )\cdot \mathbf{p}(t)$, where $\beta$ is the natural frequency of the oscillator, and $\mathbf{Z}(\phi )$ is the sensitivity function with $\mathbf{Z}(\phi )=\mathbf{Z}(\phi +2\pi )$ \cite{rf:Kura1984}.
Based on this general phase equation, we suppose the phase motion of glacial cycles as follows:
\begin{equation}
\dot{\phi} =\beta -\alpha V'(\phi )[1+\gamma F(t)], \label{eq:qp}
\end{equation}
where $V'(\phi )=\frac{dV(\phi )}{d\phi}$ is the gradient of the ice volume, $F(t)$ is a forcing function such as $F_A(t)$, and the parameters $\alpha $, $\beta$, and $\gamma$ are assumed to be positive. The specific term, $-\alpha V'(\phi )[1+\gamma F(t)]$, provides a forcing to move toward an ice volume minimum for a large astronomical forcing with $F(t)>-1/\gamma $ and to move toward an ice volume maximum for a small astronomical forcing with $F(t)<-1/\gamma $.
We can also write Eq.~(\ref{eq:qp}) as $\dot{\phi}=-\frac{\partial U(\phi, t)}{\partial \phi}$
using a time-dependent potential $U(\phi ,t)=-\beta \phi +\alpha V(\phi )[1+\gamma F(t)]$ (Fig.~\ref{fig:potential_paper}(a)). In this paper, we restrict ourselves to the simple case $|\delta |<1/4$, where the number of local minima of $U(\phi ,t)$ as a function of $\phi $ is at most one.

Let us first consider the case of constant forcing $F(t)=F_c$ is considered.
The bifurcation diagram of Eq.~(\ref{eq:qp}) for $F(t)=F_c$ is shown in Fig~\ref{fig:potential_paper}(b).
For low insolation $F_c<-\frac{1}{\gamma }\left(\frac{\beta}{\alpha (1+\delta )}+1\right)$, 
a stable equilibrium point, or a stable {\it node}, exists near the ice volume maximum $\phi _{\max}$, and 
an unstable equilibrium point, or a {\it saddle}, near the minimum $\phi _{\min}$.
Thus, the system approaches a stable glacial state. For $-\frac{1}{\gamma }\left(\frac{\beta}{\alpha (1+\delta )}+1\right)<F_c<\frac{1}{\gamma }\left(\frac{\beta}{\alpha (1-\delta )}-1\right)$, there is no equilibrium point, and glacial-interglacial oscillations emerge.
For high insolation $F_c>\frac{1}{\gamma }\left(\frac{\beta}{\alpha (1-\delta )}-1\right)$, a stable node exists near the ice volume minimum $\phi _{\min}$, and a saddle near the maximum $\phi _{\max}$.
Thus, the system approaches a stable interglacial state.

\begin{figure}[t]
\begin{center} 
\includegraphics[scale=0.3,angle=270]{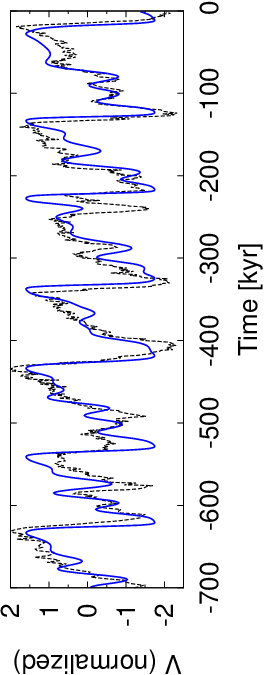}
\caption{(Color) Global ice volume reconstructed from benthic oxygen isotopic ratio (dashed line) \cite{rf:Lisi2005} and the simulated ice volume $V$ (blue line) using Eqs.~(\ref{eq:ins}) and (\ref{eq:qp}). $\alpha =1.0$, $\beta =1.0006$, $\gamma =1.0$, and $\delta =0.24$. Both curves are normalized.
} \label{fig:traj}
\end{center}
\end{figure}

Next, the case of astronomical forcing $F(t)=F_A(t)$ is studied. Unless otherwise noted, differential equations are integrated by the Runge--Kutta 4th order method with a step size of 100~yr (namely $h=0.01$). Figure~\ref{fig:traj} shows a simulated ice volume trajectory $V(\phi (t))$ for $\alpha =\gamma =1.0$, $\beta=1.0006$, and $\delta =0.24$ with initial condition $\phi (t_0)=0$ at $t_0=-20$~Myr. Here, assuming $\alpha =\gamma =1.0$ and $\delta =0.24$, the value of $\beta$ was tuned to maximize the Pearson's correlation coefficient $r$ over the past 700~kyr between the ice volume reconstruction \cite{rf:Lisi2005} and a simulated ice volume trajectory $V(\phi (t))$ with $\phi (t_0)=0$ at $t_0=-20$~Myr. This yields $r=0.76$ for $\beta=1.0006$.

\begin{figure}[t]
\begin{center} 
\includegraphics[scale=0.5,angle=270]{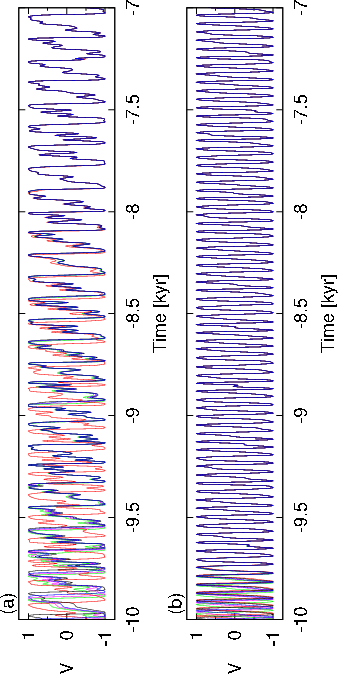}
\caption{(Color) Synchronization of ice volume trajectories $V(\phi (t))$ started from different initial conditions $\phi (t_0)= 2\pi k/5$ ($k=0,\,1,\,2,\,3,\,4$) at time $t_0=-10$~Myr under the astronomical forcing $F_A(t)$. (a) The case of 100-kyr cycles for $\alpha =1.0$, $\beta =1.0006$, $\gamma =1.0$, and $\delta =0.24$. (b) The case of 41-kyr cycles for $\alpha =0.80$, $\beta =1.65$, $\gamma =1.0$, and $\delta =0.24$.
} \label{fig:traj_convergence}
\end{center}
\end{figure}

Generally, trajectories $\phi (t)$ can depend on initial conditions $\phi (t_0)$. However, due to the astronomical forcing $F_A(t)$, trajectories starting from different initial phases $\phi (t_0)$ can converge to a single or some pieces of trajectories after a certain transient time, as shown in Figs.~\ref{fig:traj_convergence}(a) and \ref{fig:traj_convergence}(b). The transient time can last for several million years for the case of 100-kyr cycles, but it is only several cycles for the case of 41-kyr cycles.

It should be noted that the asymmetric sawtooth pattern of the ice volume trajectory $V(\phi (t))$ is not due to the subtle asymmetry of function $V(\phi)$ caused by the second harmonic $-(\delta /2)\sin 2\phi$ but due to the specific form of Eq.~(\ref{eq:qp});
the phase $\phi(t)$ increases slowly in glaciation phases and quickly in deglaciation phases, on average. The qualitative features of the following bifurcation analysis do not change even if the second harmonic is omitted or changed in the range $|\delta |<1/4$, but we keep the term because it increases the agreement with data, for example, by about 10\% from $r=0.64$ for $\delta =0$ to $r=0.76$ for $\delta =0.24$.

\section{Bifurcation analysis}\label{Sec3}
We study bifurcations of the model for the astronomical forcing $F_A(t)$ and an ideal two-frequency quasiperiodic forcing described below. We set $\gamma =1.0$ and $\delta =0.24$, and regard $\alpha$ and $\beta $ as control parameters.

\subsection{The case of the ideal two-frequency quasiperiodic forcing}
\begin{figure*}[t]
 \begin{minipage}{0.31\hsize}
  \begin{center}
   \includegraphics[scale=0.19,angle=270]{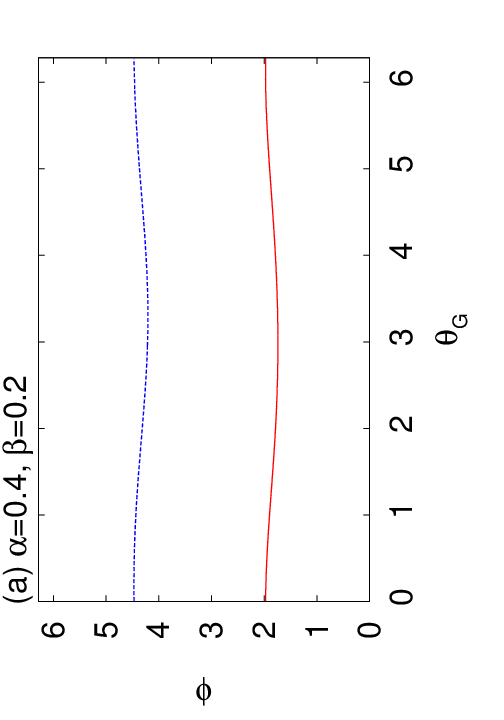}
  \end{center}
 \end{minipage}
 \begin{minipage}{0.31\hsize}
  \begin{center}
  \includegraphics[scale=0.19,angle=270]{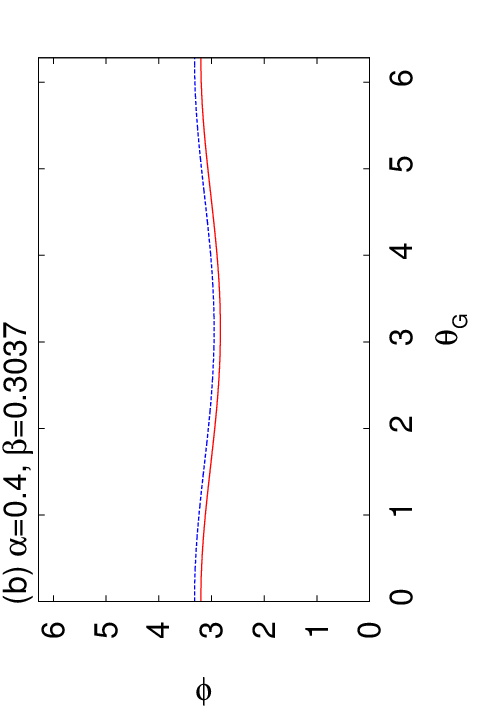}
  \end{center}
 \end{minipage}
 \begin{minipage}{0.31\hsize}
  \begin{center}
  \includegraphics[scale=0.19,angle=270]{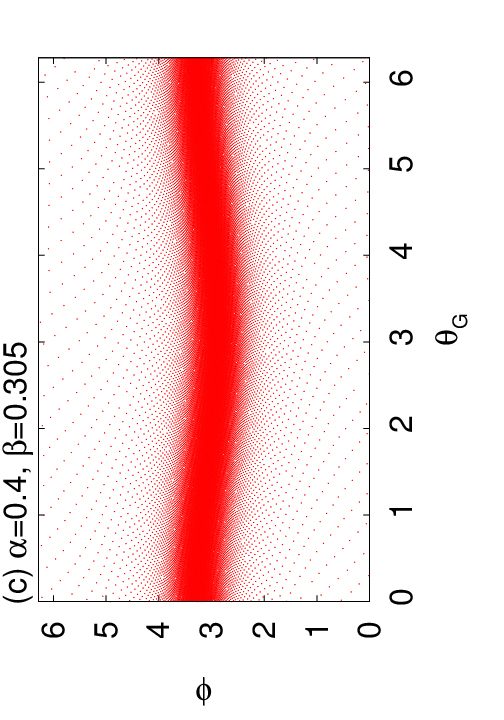}
  \end{center}
 \end{minipage}
 \begin{minipage}{0.31\hsize}
  \begin{center}
   \includegraphics[scale=0.19,angle=270]{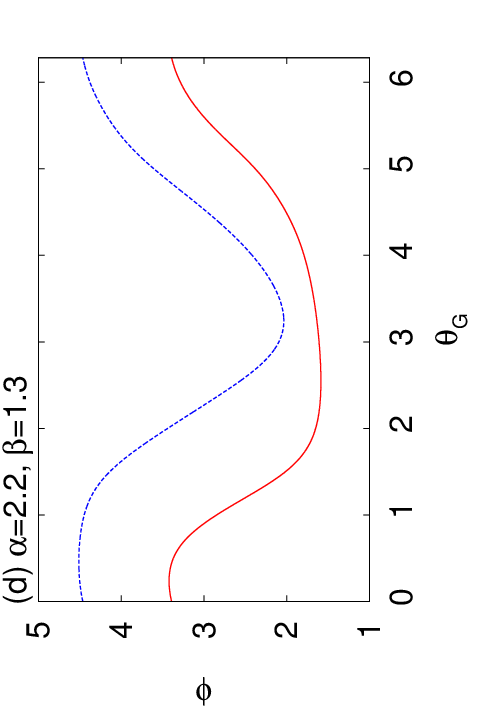}
  \end{center}
 \end{minipage}
 \begin{minipage}{0.31\hsize}
  \begin{center}
  \includegraphics[scale=0.19,angle=270]{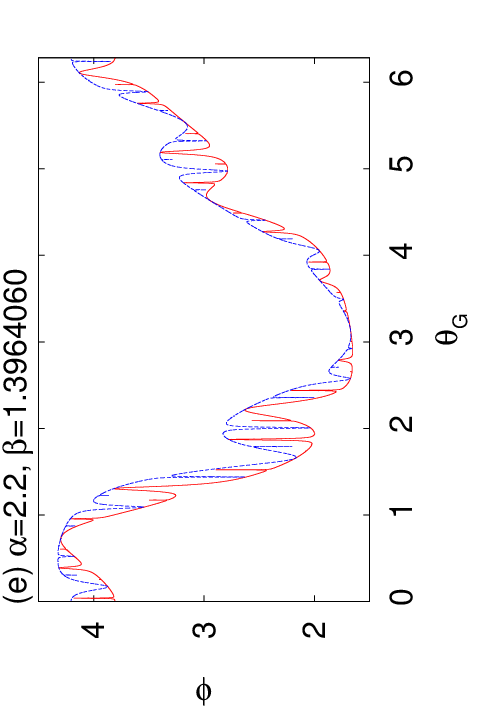}
  \end{center}
 \end{minipage}
 \begin{minipage}{0.31\hsize}
  \begin{center}
  \includegraphics[scale=0.19,angle=270]{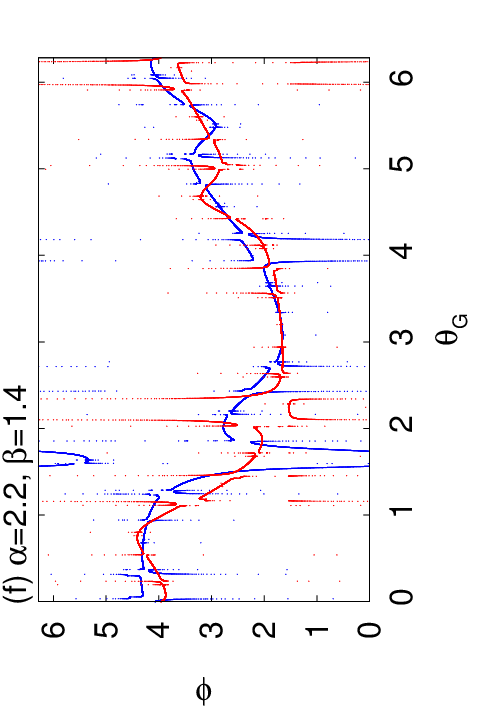}
  \end{center}
 \end{minipage}
 \caption{(Color) Poincar\'e surface of section plots at a certain constant value of $\theta _1 (\approx 2.168184)$ obtained by integrating Eq.~(\ref{eq:skew}) forward (red) and backward (blue) for a long time, respectively. Each panel corresponds to different values of parameters $\alpha$ and $\beta$: (a) in two-frequency quasiperiodic regime, (b) same but just before the smooth saddle-node bifurcation occurring at $\alpha=0.4$, $\beta_c\approx 0.303754$, (c) in three-frequency quasiperiodic regime, (d) in two-frequency quasiperiodic regime, (e) same but just before the nonsmooth saddle-node bifurcation occurring at $\alpha =2.2$, $\beta _c\approx 1.3964061$, and (f) in the SNA regime.}
\label{fig:att}
\end{figure*}

We start from the two-frequency quasiperiodic forcing 
\begin{equation}
F_G(t)=\sin \omega _1t+\sin \omega _Gt, \label{eq:ins2}
\end{equation} 
where $\omega _1$ is the most prominent frequency of the climatic precession ($2\pi /\omega _1\approx 23.7$~kyr), and $\omega _G=\omega _1 \times (\sqrt{5}-1)/2$ ($2\pi /\omega _G\approx 38.4$~kyr) is a frequency near the dominant frequency of obliquity change $2\pi /\omega _4\approx 41.0$~kyr \cite{rf:Berg1978}, which was chosen to be most incommensurate with $\omega _1$. 
Defining phase variables, $\theta _1=\omega _1t \ (\mbox{mod} \ 2\pi)$ and $\theta _G=\omega _Gt \ (\mbox{mod} \ 2\pi)$, the model can be expressed in the skew-product form
\begin{equation}
\begin{split}
&\dot{\phi}=\beta +\alpha (\cos \phi +\delta \cos 2\phi )[1+\gamma F_G(\theta _1,\,\theta _G)], \\
&\dot{\theta _1}=\omega _1, \ \ \ \ \dot{\theta _G}=\omega _G, \label{eq:skew}
\end{split}
\end{equation}
where $F_G(t)$ is redefined as $F_G(\theta _1,\,\theta _G)=\sin \theta _1+\sin \theta _G$, and the phase space is $\mathbb{T}^3$.
Similar pendulum-type systems but with additive quasiperiodic forcing have been studied in \cite{rf:Bond1985,rf:Rome1987,rf:Neum2002,rf:Kuzn2003}. The attractor of the system is obtained by solving Eq.~(\ref{eq:skew}) forward for a long time.
In the particular case of this model, the unstable invariant set of the system is found by solving Eq.~(\ref{eq:skew}) backward for a long time (this is called the {\it repeller}) \cite{rf:Neum2002}.
We plot the attractor and the repeller at a Poincar\'e surface of section (PS)
at a certain constant value of $\theta _1 (\approx 2.168184)$.\footnote{In numerical calculations for Fig.~\ref{fig:att}, a step size $h=2\pi /\omega _1 \times 0.002$ was used so that $\theta _1$ takes a constant phase in every 500 steps.}

The system~(\ref{eq:skew}) has three Lyapunov exponents: one is the Lyapunov exponent $\lambda _{\phi}$ in the direction of $\phi$, and the other two are trivially zero, corresponding to $\dot{\theta _1}=\omega _1$ and $\dot{\theta _G}=\omega _G$. The dynamics of Eq.~(\ref{eq:skew}) are classified by the Lyapunov exponent $\lambda _{\phi}$ and the winding number $W$. The winding number $W$ is defined as $$W=\lim _{t\to {\infty}}\frac{\phi (t)-\phi (0)}{t}$$ for the same Eq.~(\ref{eq:skew}) but with $\phi$ on the line $\mathbb{R}$. 
The dynamics are referred to as {\it mode-locked} if the winding number $W$ is constant with respect to slight changes of $\beta$ \cite{rf:Bjer2008}.
When the dynamics are mode-locked, the model has a two-frequency quasiperiodic attractor and a two-frequency quasiperiodic repeller, which are represented by a stable and an unstable invariant curve in PS, respectively (Fig.~\ref{fig:att}(a) for example), and 
the winding number $W$ is rationally related to the forcing frequencies, i.e., $W=(k/m)\omega _1+ (l/m)\omega _G$ ($k,l\in \mathbb{Z}$, $m\in \mathbb{N}$) \cite{rf:Bjer2008}. The denominator $m$ corresponds to the multiplicity of each invariant curve in the direction of $\phi$ \cite{rf:Bond1985}. The two-frequency quasiperiodic attractors are characterized by the negative Lyapunov exponent $\lambda _{\phi}<0$. When the dynamics are non-mode-locked, we find three-frequency quasiperiodic motion when $\lambda _{\phi}=0$ and SNAs when $\lambda _{\phi}<0$, as is the case of the general quasiperiodically forced circle map \cite{rf:Feud1997} (see Figs.~\ref{fig:att}(c) and \ref{fig:att}(f)).

The regime diagram for $F_G(t)$ is shown in Fig.~\ref{fig:bif1}(a), which is obtained by taking a grid size of $\Delta \alpha =0.01$ and $\Delta \beta =2.5\times 10^{-3}.$ The dynamics are numerically regarded as mode-locked if the change of $W$ is less than $10^{-4}$ when $\beta$ is changed either by $\pm \Delta \beta$ (white regions). The non-mode-locked regions are divided into regions with $|\lambda _{\phi}|< 5.0\times 10^{-4}$ (green) and regions with $\lambda _{\phi} \le -5.0\times 10^{-4}$ (magenta). Under these numerical criteria, we find typically the two-frequency quasiperiodic attractors in mode-locked regions (white), three-frequency quasiperiodic motion in non-mode-locked regions with $|\lambda _{\phi}|< 5.0\times 10^{-4}$ (green), and SNAs in non-mode-locked regions with $\lambda _{\phi}< -5.0\times 10^{-4}$ (magenta).  The boundaries of each regime depend on the grid size and thresholds slightly,\footnote{For the ($N+1$)-frequency quasiperiodic motion with the Lyapunov exponent $\lambda _{\phi}$ of zero, the order of the computed Lyapunov exponent converges to zero inversely proportionally to the averaging time $T$ of the local Lyapunov exponent. Typically, we have $|\lambda _{\phi}|\sim O(10^{-5})$ for $T=10^{5}$. Thus, the threshold at $\lambda _{\phi} = -5.0\times 10^{-4}$ is effective to distinguish the regimes with and without the Lyapunov exponent $\lambda _{\phi}$ of zero. The changes in the threshold between $-10^{-5}<\lambda _{\phi}<-10^{-4}$ shift the boundaries between ($N+1$)-frequency quasiperiodic motion (green) and SNA (magenta) in the direction parallel to nearby mode-locked regions. This uncertainty of the boundaries is typically less than 0.1 with respect to $\alpha $ and $\beta $. The choice of the threshold for the winding number $W$ is also based on the numerical convergence speed. The above choice gives consistent results with the analysis of the phase sensitivity exponent shown in Appendix.} but the overall structure is fairly robust.

The strangeness of attractor in the non-mode-locked regions with negative Lyapunov exponent (magenta) may also be assessed with the phase sensitivity exponent \cite{rf:Piko1995}, which characterizes the sensitivity of $\phi $ with respect to the changes in $\theta _1$ or $\theta _G$. 
The phase sensitivity exponent is computed in the Appendix, and the results show that the attractor is smooth in the mode-locked regions, and strange in the non-mode-locked regions with negative Lyapunov exponent.

The mode-locked regions form Arnol'd tongue-like structures in the parameter plane, but the width of each mode-locked region first increases and then decreases to values close to zero. Such leaf-like Arnol'd tongues are characteristic of mode-locking in quasiperiodically forced systems \cite{rf:Feud1995}, and they have been found in several models of glacial cycles \cite{rf:Saed2013,rf:Cruc2013}. 

The system exhibits two types of bifurcations on the boundaries of mode-locked regions. 
The {\it smooth saddle-node bifurcation of tori} occurs on the boundaries where the Lyapunov exponent $\lambda _{\phi}$ changes from negative to zero (between white and green regions). The {\it nonsmooth saddle-node bifurcation of tori} \cite{rf:Feud1997,rf:Feud1995,rf:Jage2009} occurs on the boundaries where $\lambda _{\phi}$ remains negative (between white and magenta regions). Here, ``tori" means the quasiperiodic attractor and repeller in the mode-locked regions (In higher-dimensional systems, the latter need not be a repeller but an unstable quasiperiodic torus of saddle type).
In this paper, we focus on the bifurcations for the mode-locked region with $W=0$,
but they occur in the other mode-locked regions as well in the same manner.

In the smooth saddle-node bifurcation, the distance between stable and unstable invariant curves decreases to zero for every value of $\theta _G$, as shown in Fig.~\ref{fig:att}(b). At the bifurcation point $\beta =\beta _c$, the stable and unstable invariant curves merge at every point in $\theta_G$ yielding three-frequency quasiperiodic motion, as shown in Fig.~\ref{fig:att}(c).
Specifically, define the distance $d (\theta _1, \theta _G)$ between a point on attractor and a point on the repeller at the same ($\theta _1,\theta _G$), and consider its maximum $d_{\max}$ and its minimum $d_{\min}$ on $(\theta _1,\theta _G)\in \mathbb{T}^2$ \cite{rf:Haro2006}.   
As shown in Fig.~\ref{fig:snbif_ch}(a), they obey a scaling law $d_{\min} \simeq C_1|\beta _c-\beta |^{0.5}$ and $d_{\max} \simeq C_2|\beta _c-\beta |^{0.5}$ ($C_1\le C_2$).
This scaling clearly holds with $C_1=C_2$ in the small forcing limit $\gamma \to 0$, where the smooth saddle-node bifurcation of tori degenerates into that of equilibrium points.

On the other hand, in the non-smooth saddle-node bifurcation, the collision of the stable invariant curves and unstable invariant curves occurs only in a countable dense set of $\theta _G$, as demonstrated in Fig.~\ref{fig:att}(e), and it creates an SNA and a strange repeller. An SNA and a strange repeller for $\alpha = 2.2$ and $\beta =1.4$ are shown in Fig.~\ref{fig:att}(f).
At the bifurcation point $\beta =\beta _c$, the minimal distance $d_{\min}$ decreases to zero, but the maximal distance $d_{\max}$ is strictly positive, as shown in Fig.~\ref{fig:snbif_ch}(b) \cite{rf:Haro2006}. 
In the nonsmooth saddle-node bifurcation, the distances do not show the scaling low $|\beta _c-\beta|^{0.5}$, which appears in the smooth saddle-node bifurcation.

\subsection{The case of the astronomical forcing}
We now consider the case of the astronomical forcing $F_A(t)$, where Eq.~(\ref{eq:skew}) is defined using $F_A(\theta _1,...,\theta _{35})=\frac{1}{a}\sum _{i=1}^{35} (s_i\sin \theta _i +c_i\cos \theta _i)$, $\dot \theta _i=\omega _i$ instead of $F_G(\theta _1,\theta _G)$. The regime diagram for $F_A(t)$ is shown in Fig.~\ref{fig:bif1}(b), which is produced in the same manner as Fig.~\ref{fig:bif1}(a).
As a natural extension from two-frequency forcing, the system exhibits a 35-frequency quasiperiodic attractor in each mode-locked region (white), 36-frequency quasiperiodic motion in non-mode-locked regions with $\lambda _{\phi} =0$ (green), and an SNA in non-mode-locked regions with $\lambda _{\phi} <0$ (magenta). The strangeness of attractor in non-mode-locked regions with negative Lyapunov exponent is assessed by the phase sensitivity exponent \cite{rf:Feud1995} in Appendix, which numerically shows that the attractor is smooth in the mode-locked regions and strange in the non-mode-locked regions with negative Lyapunov exponent.

As is the case of two-frequency forcing, smooth and nonsmooth saddle-node bifurcations occur at the boundaries of mode-locked regions.
Due to high dimensionality, the attractor and repeller are not visible but we can calculate the maximal distance $d_{\max}$ and the minimal distance $d_{\min}$ between the attractor and the repeller. 
On the boundaries with zero Lyapunov exponent, the smooth saddle-node bifurcation occurs creating 36-frequency quasiperiodic motion, with the scaling law $d_{\min} \simeq C_1|\beta _c-\beta |^{0.5}$ and $d_{\max} \simeq C_2|\beta _c-\beta |^{0.5}$, as shown in Fig.~\ref{fig:snbif_ch}(c).
On the boundaries with negative Lyapunov exponent, the nonsmooth saddle-node bifurcation occurs creating an SNA and a strange repeller. 
At the bifurcation point $\beta =\beta _c$, the minimal distance $d_{\min}$ decreases to zero, but the maximal distance $d_{\max}$ is strictly positive, as shown in Fig.~\ref{fig:snbif_ch}(d). In the nonsmooth saddle-node bifurcation, the distances do not show the scaling low $|\beta _c-\beta|^{0.5}$, which appears in the smooth saddle-node bifurcation.

\section{Discussions}
\subsection{Parameter sensitivity}
A number of previous studies report parameter sensitivity of the sequence of glacial cycles generated by simulation models \cite{rf:Pail2015,rf:Mits2014,rf:Cruc2013,rf:Weer1976,rf:Oerl1982,rf:Hogg2008,rf:Gano2011,rf:Daru2014}. In some models, parameter sensitivity is attributed to thresholds with discontinuity \cite{rf:Pail2015}, and in other models, it is attributed to a chaotic property, that is, the exponential sensitivity to initial conditions \cite{rf:Daru2014}. However, parameter sensitivity can appear even when the models are smooth dynamical systems and nonchaotic. An example is Saltzman-Maasch (1990) model \cite{rf:Salt1990} as pointed out in \cite{rf:Cruc2013}. Our phase oscillator model also shows such a parameter sensitivity.  

To indicate ``the region of 100~kyr cycles" in parameter space, we again calculate the Pearson's correlation coefficient $r$ over the past 700~kyr between the ice volume reconstruction \cite{rf:Lisi2005} and a simulated ice volume trajectory with $\phi (t_0)=0$ at $t_0=-20$~Myr. 
The parameter points with high correlation $r>0.7$ are plotted by symbol ``+'' in Fig.~\ref{fig:bif1}(b).\footnote{High values of correlation coefficient $r$ with $r>0.7$ can be obtained in regions with 36-frequency quasiperiodic motion (green) or in mode-locked regions with $m\geq 2$ (white) by chance because trajectories depend on initial conditions over the past 700~kyr.} 
They are distributed in a narrow region, where the winding number $W$ is constrained between 0.58 and 0.66.
Around the region of 100~kyr cycles, the mode-locked states are unlikely to occur because they have little measure. Even if the mode-locked states occurred, they would be fragile against small parameter changes.

\begin{figure}[t]
\begin{center} 
\includegraphics[scale=0.5,angle=270]{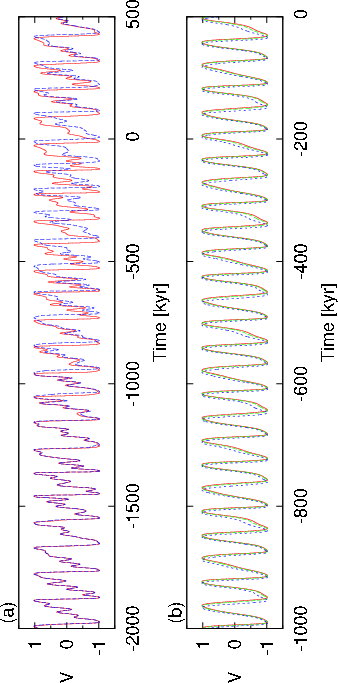}
\caption{(Color) (a) Simulated ice volume trajectories for two slightly different values in SNA regime: $\beta=1.0006$ (red solid line) and $\beta=1.0001$ (blue dashed line) ($\alpha =1.0$, $\gamma =1$, and $\delta =0.24$). (b) Simulated ice volume trajectories for three different values in 41-kyr mode-locked region : $\beta=1.61$ (red solid line), $\beta=1.65$ (green dashed), and $\beta =1.69$ (blue dotted line) ($\alpha =0.8$, $\gamma =1$, and $\delta =0.24$).
} \label{fig:sensitivity}
\end{center}
\end{figure}

If the 100~kyr cycles are non-mode-locked, the dynamics of SNA or the dynamics of ($N+1$)-frequency quasiperiodic motion with the Lyapunov exponent of zero are candidates for the 100~kyr cycles. The dynamics of SNAs show the synchronizing property under the same astronomical forcing, as shown in Fig.~\ref{fig:traj_convergence}(a), which is often considered as an important feature for the models of glacial cycles. However, in the SNA regime, the sequence of glacial cycles can be highly sensitive to parameter changes. Two simulated ice volume trajectories for two slightly different parameter values $\beta=1.0006$ (red) and $\beta=1.0001$ (blue) are shown in Fig.~\ref{fig:sensitivity}(a). The deviation of the trajectories with nearby parameters is reminiscent of the well-known butterfly effect in chaotic systems but this system is nonchaotic and even has a negative Lyapunov exponent $\lambda _\phi$ for these parameter values of $\beta $. Since SNAs have strange geometrical structures as shown in Fig.~\ref{fig:att}(f), subtle changes of parameter values can induce large shifts of the points in the attractors.

On the other hand, the 41~kyr cycles is caused by the mode-locking to the 41~kyr component of astronomical forcing (due to obliquity motion) in this model. The mode-locking region of the 41~kyr cycles is relatively wide (see the white region labeled by $W=\omega _4$ in Fig.~\ref{fig:bif1}(b)). Thus, this mode-locking is likely to occur. This is consistent with the previous study \cite{rf:Ashk2004}, which proposes that the phase of 41~kyr cycles is locked by the astronomical forcing.

The sequence of the mode-locked 41~kyr cycles is relatively robust to parameter changes. Figure~\ref{fig:sensitivity}(b) shows ice volume trajectories corresponding to different values of $\beta $ in the mode-locked region: $\beta=1.61$, $\beta=1.65$, and $\beta =1.69$. The other parameters are $\alpha =0.8$, $\gamma =1$, and $\delta =0.24$.

\subsection{Middle Pleistocene transition}
\begin{figure}[h]
\begin{center} 
\includegraphics[scale=0.55,angle=270]{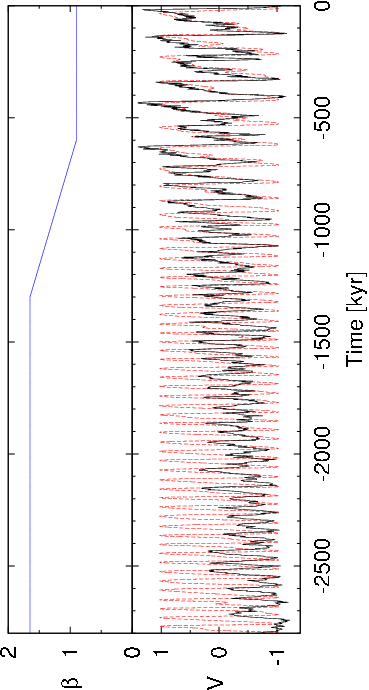}
\caption{(Color) Glacial cycles over the Quaternary: The global ice volume reconstructed from the oxygen isotope ratio (black solid line) \cite{rf:Lisi2005} and the simulated ice volume $V(\phi (t))$ (red dashed line). The former is scaled as $1.3\times \delta ^{18}O-4$ for comparison. The parameter $\beta $ is fixed at $\beta =1.65$ until $-1300$~kyr and fixed at $\beta =0.9$ after $-600$~kyr, between which $\beta $ is decreased linearly interpolating the two values. The other parameters are $\alpha =0.8$, $\gamma =1.0$, and $\delta =0.24$. The sequence is robust to changes in initial conditions because of the strong synchronizing property of the 41-kyr cycles before the middle-Pleistocene transition.
} \label{fig:MPT_bb}
\end{center}
\end{figure}

The middle Pleistocene transition (MPT) began $\sim $1.2 Myr ago and was complete by $\sim $0.7 Myr ago, through which the average period of glacial cycles changed from $\sim $40~kyr to $\sim $100~kyr, accompanying an increase of amplitude \cite{rf:Clar2006}. 
A full investigation of the causes and dynamics of the MPT is beyond the scope of this paper. 
In particular, a phase model provides no information on amplitude. Yet, we note that a frequency change similar to the one observed during the MPT is obtained when the parameter $\beta$ is changed over the period of MPT, as shown in Fig.~\ref{fig:MPT_bb}. The parameter $\beta $ is fixed at $\beta =1.65$ in the model-locked region with $W=\omega _4$ until $1300$~kyr ago (see Fig.~\ref{fig:bif1}(b)). Then, $\beta $ is decreased linearly until it reaches $\beta =0.9$ at $600$~kyr ago. After $600$~kyr ago, $\beta $ is kept at $\beta =0.9$. The other parameters are set at $\alpha =0.8$, $\gamma =1.0$, and $\delta =0.24$. The sequence of the glacial cycles in Fig.~\ref{fig:MPT_bb} is robust to changes in initial conditions because of the strong synchronizing property of the 41-kyr cycles before the MPT (cf. Fig.~\ref{fig:traj_convergence}(b)).    

\section{Summary}\label{Sec4}
We introduced a phase oscillator model of glacial cycles and analyzed the
bifurcations of the model for the ideal two-frequency quasiperiodic forcing
and for the astronomical forcing. It was shown that SNAs appear through
nonsmooth saddle-node bifurcations of tori in the model.
Based on the results for the phase oscillator model, we conjecture that the bifurcations from quasiperiodic attractors to SNAs found in oscillator models of glacial cycles \cite{rf:Cruc2013} are also nonsmooth saddle-node bifurcations.
The regime diagram in Fig.~\ref{fig:bif1}(b) indicates that mode-locking is likely to occur for the 41~kyr glacial cycles but not likely for the 100~kyr glacial cycles. The sequence of mode-locked 41~kyr cycles is robust to small parameter changes. However, the sequence of 100~kyr glacial cycles can be sensitive to parameter changes when the system has an SNA. 

Long transient dynamics of million-year scale can be observed for the 100-kyr glacial cycles though it can vanish quickly for the 41-kyr cycles (Figs.~\ref{fig:traj_convergence}(a) and \ref{fig:traj_convergence}(b)). Given that the Quaternary ice age is $\sim$3~Myr, and that there can be perturbations to the system, transient trajectories may be important to understand the glacial cycles, not only just on the attractor. This problem will be explored elsewhere \cite{rf:Mits2015}. 

\section*{Acknowledgments}
We would like to thank the two anonymous reviewers for their suggestions and comments. TM and KA are supported by the Aihara Project, the FIRST program from JSPS, initiated by CSTP. TM is also partly supported by the ITOP project, ERC-StG grant 239604. MC is supported by the ITOP project ERC-StG grant 239604 and by the Belgian National Fund of Scientific Research.

\section*{Appendix. Strangeness of attractors in the non-mode-locked regions with the negative Lyapunov exponent}
\begin{figure*}[t]
 \begin{minipage}{0.49\hsize}
  \begin{center}
   \includegraphics[scale=0.45,angle=270]{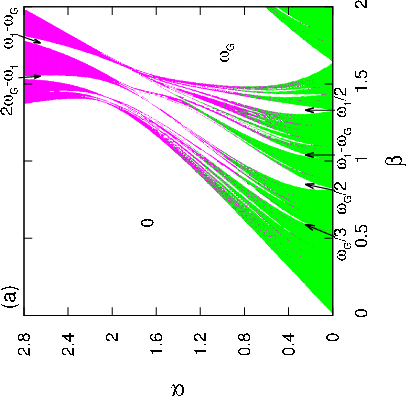}
  \end{center}
 \end{minipage}
 \begin{minipage}{0.49\hsize}
  \begin{center}
  \includegraphics[scale=0.45,angle=270]{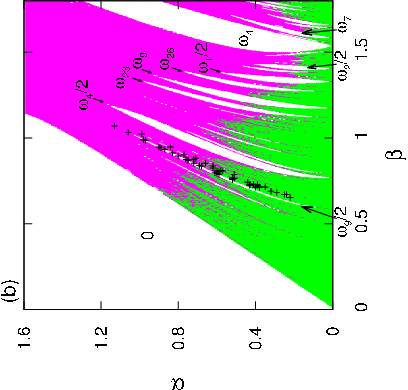}
  \end{center}
 \end{minipage}
 \caption{(Color) (a) Regime diagram for the two-frequency forcing $F_G(t)$ ($N=2$) and (b) regime diagram for the astronomical forcing $F_A(t)$ ($N=35$): the mode-locked regions corresponding to $N$-frequency quasiperiodic attractors (invariant curves in PS) (white), the regions with $|\lambda _{\phi}|< 5.0\times 10^{-4}$ typically corresponding to ($N+1$)-frequency quasiperiodic motion (green), and the regions with $\lambda _{\phi} \le -5.0\times 10^{-4}$ typically
corresponding to SNAs (magenta). Prominent mode-locked regions are shown with their rotation numbers. The parameter points corresponding to correlation coefficient $r$ larger than $0.7$ are marked by symbol ``+" (see text for the details of the calculation of $r$).}
 \label{fig:bif1}
\end{figure*}

We present numerical support for the existence of SNAs in the non-mode-locked regions with the negative Lyapunov exponent $\lambda _{\phi }<0$ (magenta regions of Figs.~\ref{fig:bif1}(a) and (b)).
Denote formally the phase equation of an oscillator under an $N$-frequency forcing by
\begin{equation}
\left\{
\begin{split} 
\dot{\mathbf{\theta }}&=\mathbf{\omega }, \\
\dot{\phi }&=f(\phi,\mathbf{\theta }),
\end{split} \label{eq:A1}
\right.
\end{equation}
where $\mathbf{\theta }=(\theta _1, \theta _2, ..., \theta _N)^\mathrm{T}$ and $\mathbf{\omega }=(\omega _1, \omega _2, ..., \omega _N)^\mathrm{T}$.
When the system~(\ref{eq:A1}) has an $N$-frequency quasiperiodic attractor, it is represented by a certain smooth function $\phi =H(\theta )$. On the other hand, when the system~(\ref{eq:A1}) has an SNA, the relationship between $\theta $ and $\phi $ is discontinuous. 

The phase sensitivity function $\Gamma (t)$ of variable $\phi $ with respect to variable $\theta _i$ is defined as 
\begin{equation}
\Gamma (t)=\min _{\{\phi (0),\,\theta (0)\}}\left\{ \max _{0\leq \tau \leq t} \left| \frac{\partial \phi (\tau )}{\partial \theta _i}\right| \right\}.
\end{equation}
This derivative $\partial \phi (\tau )/\partial \theta _i$ is obtained by integrating 
\begin{equation}
\frac{d}{dt}\left( \frac{\partial \phi}{\partial \theta _i}\right)=\frac{\partial f(\phi,\,\theta )}{\partial \phi}\left( \frac{\partial \phi}{\partial \theta _i}\right)+\frac{\partial f(\phi ,\,\theta )}{\partial \theta _i}, \label{eq:phs}
\end{equation}
along the solution of Eq.~(\ref{eq:A1}).
When the system~(\ref{eq:A1}) has an $N$-frequency quasiperiodic attractor, the state $(\theta (t),\,\phi(t))$ and the quantity $\partial \phi (t )/\partial \theta _i$, which started from some initial conditions, approach the attractor and its derivative $H'(\theta (t))$, respectively, as time $t$ increases \cite{rf:Piko1995}.
Thus, $\Gamma (t)$ is bounded for $N$-frequency quasiperiodic attractors. If we write the phase sensitivity function formally as $\Gamma (t)\simeq t^{\mu}$ for large $t$, the exponent $\mu$ is zero for $N$-frequency quasiperiodic attractors. The exponent $\mu$ is called the {\it phase sensitivity exponent}.
On the other hand, SNAs do not have finite derivatives by definition, but we can calculate the quantity $\partial \phi (t )/\partial \theta _i$ (and $\Gamma (t)$) along orbits approaching an SNA. It is known that the phase sensitivity function grows as $\Gamma (t)\simeq t^{\mu}$ with $\mu >0$ for SNAs (see \cite{rf:Piko1995,rf:Feud2006} for detail). 

\begin{figure}[t]
 \begin{minipage}{0.49\hsize}
  \begin{center}
   \includegraphics[scale=0.18,angle=270]{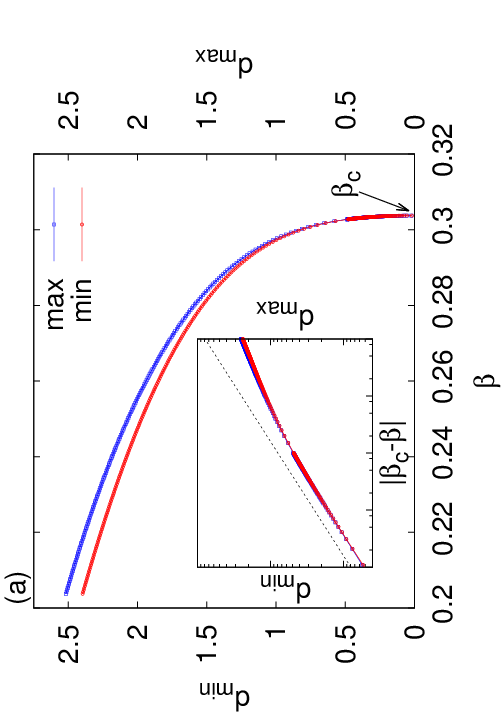}
  \end{center}
 \end{minipage}
 \begin{minipage}{0.49\hsize}
  \begin{center}
  \includegraphics[scale=0.18,angle=270]{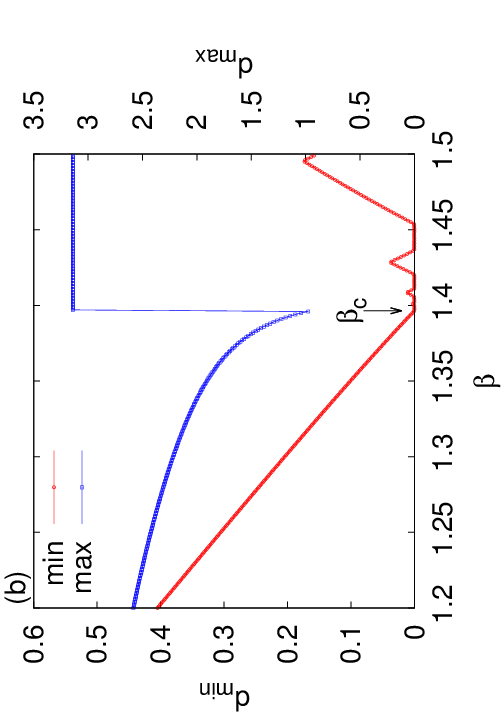}
  \end{center}
 \end{minipage}
 \begin{minipage}{0.49\hsize}
  \begin{center}
   \includegraphics[scale=0.18,angle=270]{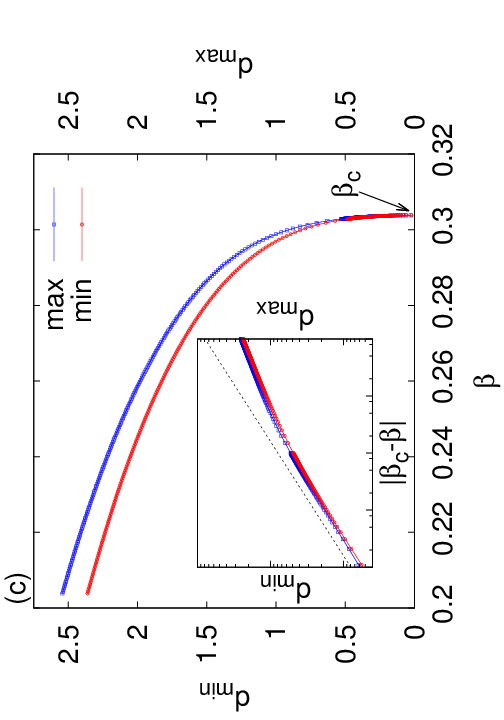}
  \end{center}
 \end{minipage}
 \begin{minipage}{0.49\hsize}
  \begin{center}
  \includegraphics[scale=0.18,angle=270]{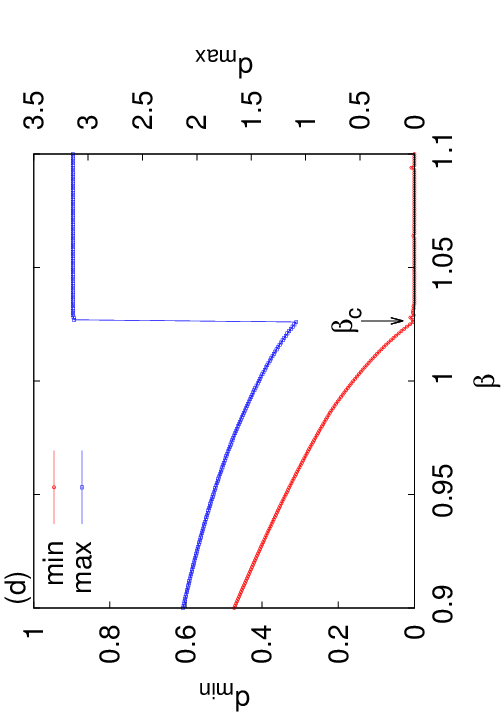}
  \end{center}
 \end{minipage}
 \caption{(color) The minimal distance $d_{\min}$ and the maximal distance $d_{\max}$ between the attractor and the repeller as functions of $\beta $: (a) a smooth saddle-node bifurcation along $\alpha =0.4$ for the two-frequency forcing $F_G(t)$. 
Inset is a log-log plot showing the scaling $\propto |\beta _c-\beta|^{0.5}$ just before the bifurcation occurring at $\beta =\beta _c$.
(b) a nonsmooth saddle-node bifurcation along $\alpha =2.2$ for $F_G(t)$. (c) a smooth saddle-node bifurcation along $\alpha =0.4$ for the astronomical forcing $F_A(t)$. (d) a nonsmooth saddle-node bifurcation along $\alpha =1.4$ for $F_A(t)$. The panels (a) and (c) show the results only for the region with attractor $\beta <\beta _c$. 
}
 \label{fig:snbif_ch}
\end{figure}

For the case of two-frequency forcing $F_G(\theta _1,\theta _G)$, we calculate the phase sensitivity function $\Gamma (t)$ with respect to $\theta _G$, changing parameter $\beta$, where the other parameters are set to $\alpha =2.2$, $\gamma =1$, and $\delta =0.24$, as in Fig.~\ref{fig:snbif_ch}(b). For the case of astronomical forcing $F_A(\theta _1,\,...,\,\theta _{35})$, the phase sensitivity function $\Gamma (t)$ is calculated with respect to $\theta _4$, changing parameter $\beta$, where the other parameters are set to $\alpha =1.4$, $\gamma =1$, and $\delta =0.24$, as in Fig.~\ref{fig:snbif_ch}(d). The phase sensitivity exponent $\mu$ is measured from the growth of the phase sensitivity function over the time interval [$10^5,\,10^7$].

\begin{figure*}[t]
 \begin{minipage}{0.49\hsize}
  \begin{center}
   \includegraphics[scale=0.4,angle=270]{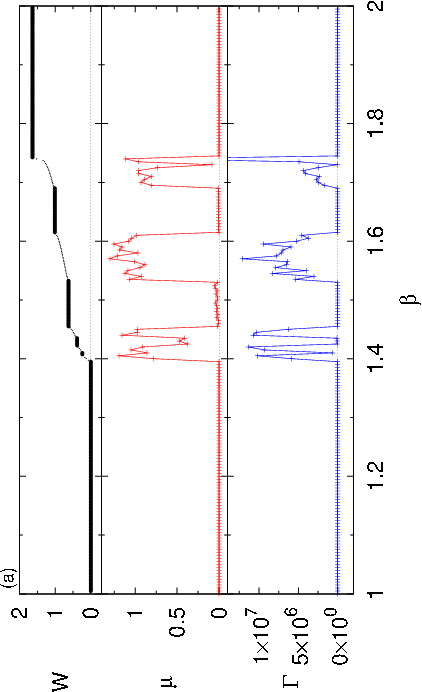}
  \end{center}
 \end{minipage}\\
 \begin{minipage}{0.49\hsize}
  \begin{center}
  \includegraphics[scale=0.4,angle=270]{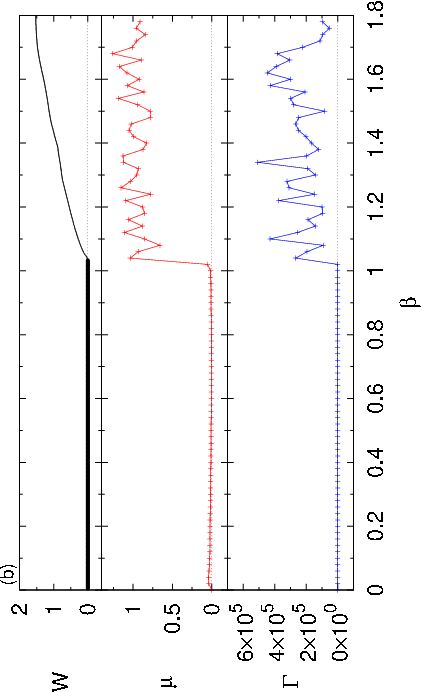}
  \end{center}
 \end{minipage}
 \caption{
The winding number $W$, the phase sensitivity exponent $\mu$, and the value of the phase sensitivity function $\Gamma$ at $t=10^7$ as functions of parameter $\beta$, respectively. (a) The case for the two-frequency forcing $F_G(t)$ with parameters $\alpha =2.2$, $\gamma =1$, and $\delta =0.24$. The phase sensitivity function is calculated with respect to $\theta _G$. (b) The case for the astronomical forcing $F_A(t)$ for with parameters $\alpha =1.0$, $\gamma =1$, and $\delta =0.24$. Mode-locked regions and non mode-locked regions are shown by thick points and thin points, respectively,  in the graph of the winding number $W$. The phase sensitivity function is calculated with respect to $\theta _4$. The exponent $\mu$ is calculated from the growth of the phase sensitivity function over the time interval [$10^5,\,10^7$]. The phase sensitivity function is minimized for 10 initial conditions. 
} \label{fig:ps}
\end{figure*}

Figures~\ref{fig:ps}(a) and \ref{fig:ps}(b) show the winding number $W$, the phase sensitivity function $\mu$, and the value of phase sensitivity function $\Gamma $ at $t=10^7$ as functions of $\beta$.
Mode-locked regions and non mode-locked regions are shown by thick points and thin points, respectively, in the graph of the winding number $W$. The phase sensitivity exponent $\mu$ takes a positive value enough away from zero in the non-mode-locked regions. This means that the attractors in the non-mode-locked regions are typically strange. On the other hand, the phase sensitivity exponent $\mu$ takes the value of zero in most of the mode-locked regions, which indicates $N$-frequency quasiperiodic attractors. In some regions ($1.4075\lesssim \beta \lesssim 1.41$ and $1.4225\lesssim \beta \lesssim 1.435$ in Fig.~\ref{fig:ps}(a)), the numerical value of $\mu$ obtained with a simulation length of $10^7$ can be positive due to slow convergence. However, the value of the phase sensitivity function $\Gamma (10^7)$ itself is small when compared to non-mode-locked regions.  



\bibliographystyle{elsarticle-num}
\bibliography{<your-bib-database>}



{}


\end{document}